\documentclass{stacs_proc}

\usepackage{enumerate}
\usepackage{url}
\usepackage{tikz}
\usetikzlibrary{shapes}
\usetikzlibrary{snakes}
\usetikzlibrary{arrows} 
\usepackage{amsmath, amsthm, amssymb}
\usepackage{mathtools}
\usepackage{stmaryrd}
\usepackage{xspace}

\newcommand{\mathLogic}[1]{\ensuremath{\mathrm{#1}}\xspace}
\newcommand{\sem}[1]{\ensuremath{\llbracket#1\rrbracket}}
\newcommand{\ie}{i.e.\ }
\newcommand{\eg}{e.g.\ }

\renewcommand{\phi}{\varphi}

\newcommand{\muc}{$\mu$-calculus\xspace}
\newcommand{\qmu}{quantitative \muc}

\newcommand{\Qmu}{\mathLogic{Q\mu}} 
\newcommand{\Lmu}{\mathLogic{L_\mu}} 
\newcommand{\QML}{\mathLogic{QML}} 

\newcommand{\LTL}{\mathLogic{LTL}} 
\newcommand{\CTL}{\mathLogic{CTL}} 
\newcommand{\CTLstar}{\mathLogic{CTL^*}} 
\newcommand{\PDL}{\mathLogic{PDL}} 

\newcommand{\Diam}{\ensuremath{\Diamond}} 

\newcommand{\semk}[1]{\ensuremath{\llbracket#1\rrbracket^{\calK}}} 
\newcommand{\semek}[1]{
  \ensuremath{\llbracket#1\rrbracket_{\varepsilon}^{\calK}}
}
\newcommand{\semsub}[1]{
  \ensuremath{\llbracket#1\rrbracket_{\varepsilon[X \leftarrow f]}^{\calK}}
}

\newcommand{\Rplusinf}{\ensuremath{\mathbb{R}^+_{\infty}}} 
\newcommand{\Rplus}{\ensuremath{\mathbb{R}^{+}}}

\newcommand{\Func}{\ensuremath{\mathcal{F}}}

\newcommand{\pzero}{Player~$0$\xspace}
\newcommand{\pone}{Player~$1$\xspace}

\newcommand{\val}{\mathLogic{val}} 
\newcommand{\mc}{\mathLogic{MC}} 
\newcommand{\Win}[1]{\mathLogic{Win}_{#1}}

\newcommand{\calG}{\ensuremath{\mathcal{G}}}

\newcommand{\calK}{\ensuremath{\mathcal{K}}}
\newcommand{\calQ}{\ensuremath{\mathcal{Q}}}
\newcommand{\calV}{\ensuremath{\mathcal{V}}}
\newcommand{\N}{\mathbb{N}} 

\newcommand{\Kk}{\ensuremath{\mathcal{K}}}

\begin{document}

\title[Model Checking Games for the Quantitative $\mu$-Calculus]
{Model Checking Games for the Quantitative $\mu$-Calculus}

\author[mgi]{D. Fischer}{Diana Fischer}
\author[mgi]{E. Gr\"adel}{Erich Gr\"adel}
\author[mgi]{\L. Kaiser}{\L{}ukasz Kaiser}

\address[mgi]{Mathematische Grundlagen der Informatik, RWTH Aachen}
\email{{fischer,graedel,kaiser}@logic.rwth-aachen.de}



\begin{abstract} 
We investigate quantitative extensions of modal logic and the
modal $\mu$-calculus, and study the question
whether the tight connection between logic and games
can be lifted from the qualitative logics to their
quantitative counterparts. 
It turns out that, if the quantitative $\mu$-calculus is defined
in an appropriate way respecting the duality
properties between the logical operators, then its model checking
problem can indeed be characterised by a quantitative variant of
parity games. However, these quantitative games have quite different
properties than their classical counterparts, in particular
they are, in general, not positionally determined. 
The correspondence between the logic and the games goes both ways:
the value of a formula on a quantitative transition system coincides
with the value of the associated quantitative game, and conversely,
the values of quantitative parity games are definable
in the quantitative $\mu$-calculus.
\end{abstract}

\maketitle

\stacsheading{2008}{301-312}{Bordeaux}
\firstpageno{301}

\section{Introduction}
There have been a number of recent proposals to extend the 
common qualitative, i.e. two-valued, logical formalisms 
for specifying the behaviour
of concurrent systems, such as propositional 
modal logic ML, the temporal logics LTL
and CTL, and the modal \muc \Lmu, to quantitative formalisms.
In quantitative logics, the formulae can take, at a given state of a 
system, not just the values \emph{true} and \emph{false}, 
but quantitative values, for instance from the 
(non-negative) real numbers.
There are several scenarios and applications
where it is desirable to replace purely qualitative statements
by quantitative ones, which can be of very different nature: 
we may be interested 
in the probability of an event, the value 
that we assign to an event may depend on how late it occurs, 
we can ask for the number of occurrences of an event in a play, and
so on. We can consider transition structures, where already 
the atomic propositions take numeric values, or we can ask about 
the `degree of satisfaction' of a property.
There are several papers that deal with either of these topics,
resulting in different specification formalisms and in different
notions of transition structures. In particular, due to the 
prominence and importance of the modal $\mu$-calculus in verification, 
there have been several attempts to define a quantitative $\mu$-calculus.
In some of these, the term quantitative refers to probability, \ie the 
logic is interpreted over probabilistic transition systems
\cite{McIver}, or used to describe winning conditions in stochastic
games \cite{AlfaroM04,Alfaro03,Hugo07}.
Other variants introduce quantities by allowing discounting in the respective
version of a ``next''-operator for qualitative transition systems 
\cite{Alfaro03}, Markov decision processes and Markov chains 
\cite{AlfaroFHMS05}, and for stochastic games \cite{AlfaroHM03}.

While there certainly is ample motivation to extend qualitative
specification formalisms to quantitative ones, there also are
problems. As has been observed in many areas of mathematics,
engineering and computer science where logical
formalisms are applied, quantitative formalisms in general lack 
the clean and clear mathematical theory of their qualitative 
counterparts, and many of the 
desirable mathematical and algorithmic properties tend to get lost.
Also, the definitions of quantitative formalisms are often ad hoc
and do not always respect the properties that are  
required for logical methodologies. In this paper we have a closer look
at quantitative modal logic and the quantitative $\mu$-calculus
in terms of their description by appropriate semantic
games. The close connection to games is
a fundamental aspect of logics. The evaluation of logical formulae
can be described by model checking games, played by two players  
on an arena which is formed as the product of a structure 
$\Kk$ and a formula $\psi$. One player (Verifier) attempts to prove 
that $\psi$ is
satisfied in $\Kk$ while the other (Falsifier) tries to refute this.

For the modal $\mu$-calculus $L_\mu$, model checking is described
by \emph{parity games}, and this connection is of crucial importance
for the model theory, the algorithmic evaluation and the
applications of the $\mu$-calculus. Indeed, most 
competitive model checking algorithms for \Lmu are based
on algorithms to solve the strategy problem in parity games \cite{spm}.
Furthermore, parity games enjoy nice properties like positional determinacy and
can be intuitively understood: often, the best way to make sense of 
a $\mu$-calculus formula is to look at the associated game. 
In the other direction, winning regions of  parity games (for any
fixed number of priorities) are definable in the modal $\mu$-calculus. 

In this paper, we explore the question to what extent the relationship
between the $\mu$-calculus and parity games can be extended to a quantitative
$\mu$-calculus and appropriate quantitative model checking games.
The extension is not straightforward, and requires that one
defines the quantitative $\mu$-calculus in the `right' way, 
so as to ensure that it has appropriate closure and duality properties 
(such as closure under negation, De Morgan equalities, quantifier
and fixed point dualities) to make it amenable to a game-based
approach. 
Once this is done, we can indeed construct a quantitative
variant of parity games, and prove that they are the
appropriate model checking games for the quantitative
$\mu$-calculus. As in the classical setting the correspondence 
goes both ways: the value of a formula in a structure coincides with
the value of the associated model checking game, and conversely,
the values of quantitative parity games (with a fixed number of priorities)
are definable in the quantitative $\mu$-calculus.
However, the mathematical properties of quantitative
parity games are different from their qualitative counterparts.
In particular, they are, in general, not positionally determined,
not even up to approximation. The proof that the quantitative
model checking games correctly describe the value of the
formulae is considerably more difficult than for the classical case. 

As in the classical case, model checking games lead to a
better understanding of the semantics and expressive power
of the quantitative $\mu$-calculus. Further, the game-based
approach also sheds light on the consequences
of different choices in the design of the quantitative
formalism, which are far less obvious than for classical logics.

\section{Quantitative $\mu$-calculus}   
In \cite{dA:metrics}, de Alfaro, Faella, and Stoelinga introduce a 
quantitative \muc,
that is interpreted over metric transition systems,
where predicates can take values in arbitrary metric spaces.
Furthermore, their \muc allows discounting in modalities
and is studied in connection with quantitative versions
of basic system relations such as bisimulation.

We base our calculus on the one proposed in \cite{dA:metrics}
but modify it in the following ways.
\begin{enumerate}
\item We decouple discounts from the modal operators.
\item We allow discount factors to be greater than one.
\item In the definition of transition systems we allow additional discounts
    on the edges.
\end{enumerate}

These changes make the logic more robust and more general,
and, as we will show in the next section, will permit us
to introduce a negation operator with the
desired duality properties that are fundamental to
a game-based analysis.

%
%

Quantitative transition systems, similar to the ones introduced
in \cite{dA:metrics} are directed graphs equipped with quantities at 
states and discounts on edges. In the sequel, $\Rplus$ is the
set of non-negative real numbers, and
$\Rplusinf:=\Rplus\cup\{\infty\}$.

\begin{definition}
A \emph{quantitative transition system (QTS)} is a tuple
\[\calK = (V, E, \delta, \{P_i\}_{i \in I}),\]
consisting of a directed graph $(V,E)$,
a discount function $\delta: E \to \Rplus\setminus\{0\}$
and functions  $P_i: V \to \Rplusinf$,
that assign to each state the values of the predicates
at that state.

A transition system is \emph{qualitative}
if all functions $P_i$ assign only the values $0$ or $\infty$,
\ie $P_i: V \to \{0, \infty\}$, where 0 stands for false 
and $\infty$ for true, and it is \emph{non-discounted} 
if $\delta(e)=1$ for all $e\in E$.
\end{definition}

We now introduce a quantitative version of the modal
\muc to describe properties of quantitative transition systems.

\begin{definition}
Given a set $\calV$ of variables $X$, predicate functions $\{P_i\}_{i \in I}$,
discount factors  $d \in \Rplus$ and constants $c \in \Rplus$,
the formulae of  \emph{quantitative \muc} (\Qmu) can be built
in the following way:

\begin{enumerate} \label{fp}
\item $|P_i - c|$ is a \Qmu-formula,
\item $X$ is a \Qmu-formula,
\item if $\phi, \psi$ are \Qmu-formulae, then so are
    $(\phi \land \psi)$ and $(\phi \lor \psi)$,
\item if $\phi$ is a \Qmu-formula, then so are $\Box \phi$ and $\Diam \phi$,
\item if $\phi$ is a \Qmu-formula, then so is $d \cdot \phi$,
\item if $\phi$ is a formula of \Qmu,
    then $\mu X.\phi$ and $\nu X.\phi$ are formulae of \Qmu. 
\end{enumerate}
\end{definition}

Formulae of \Qmu are interpreted over quantitative transition systems.
Let $\Func$ be the set of functions $f: V \to \Rplusinf$,
with $f_1 \leq f_2$ if $f_1(v) \leq f_2(v)$ for all $v$.
Then $(\Func, \leq)$ forms a complete lattice
with the constant functions $f=\infty$ as
top element and  $f=0$ as bottom element.

Given an interpretation $\varepsilon : {\calV} \to \Func$, 
a variable $X \in \calV$,
and a function $f \in \Func$, we denote by $\varepsilon[X \leftarrow f]$
the interpretation $\varepsilon'$, such that $\varepsilon'(X) = f$ and
$\varepsilon'(Y) = \varepsilon(Y)$ for all $Y \neq X$. 

\begin{definition}
Given a QTS $\calK = (V,E,\delta, \{P_i\}_{i\in I})$ 
and an interpretation $\varepsilon$,
a $\Qmu$-formula yields a valuation function $\semek{\phi} : V \to \Rplusinf$
defined as follows:

\begin{enumerate}
\item $ \semek{|P_i-c|}(v) = |P_i(v) - c|$,
\item $ \semek{\phi_{1} \land \phi_{2}} = 
        \min \{\semek{\phi_{1}},\semek{\phi_{2}}\}$ and
      $ \semek{\phi_{1} \lor \phi_{2}} = 
        \max \{\semek{\phi_{1}},\semk{\phi_{2}}\}$,
\item $ \semek{\Diam \phi}(v) = 
        \sup_{v' \in vE} \delta(v, v') \cdot \semek{\phi}(v')$ and
      $ \semek{\Box \phi}(v) = 
        \inf_{v' \in vE} \frac{1}{\delta(v,v')}\semek{\phi}(v')$,
\item $\semek{d \cdot \phi}(v) = d \cdot \semek{\phi}(v)$,
\item $\semek{X} = \varepsilon(X) $,
\item $\semek{\mu X.\phi} = \inf \{ f \in \Func : f = \semsub{\phi}\}$,
\item $\semek{\nu X.\phi} = \sup \{ f \in \Func : f = \semsub{\phi}\}$.
\end{enumerate}

For formulae without free variables, we can simply write $\semk{\phi}$
rather than $\semek{\phi}$.
\end{definition}

We call the fragment of \Qmu consisting of formulae without fixed-point
operators \emph{quantitative modal logic} \QML.
If \Qmu is interpreted over qualitative transition systems, it coincides
with the classical \muc and we say that $\calK, v$ is a model of
$\phi$, $\calK, v \models \phi$ if ${\sem \phi}^\calK(v)= \infty$.
Over non-discounted quantitative transition systems, the definition
above coincides with the one in \cite{dA:metrics}. For discounted
systems we take the natural definition for $\Diam$ and use the
dual one for $\Box$, thus the $\frac{1}{\delta}$ factor.
As we will show, this is the only definition for which
there is a well-behaved negation operator and with a close relation
to model checking games.

We always assume the formulae to be \emph{well-named},
\ie each fixed-point variable is bound only once and no 
variable appears both free and bound and we use the notions of
\emph{alternation level} and \emph{alternation depth} in 
the usual way, as defined in \eg \cite{Graedel03}.



Note that all operators in $\Qmu$ are monotone, thus guaranteeing
the existence of the least and greatest fixed points, and their inductive
definition according to the Knaster-Tarski Theorem stated below.

\begin{proposition}\label{inductivefixpoints}
The least and greatest fixed points exist and can be computed inductively:
$\semek{\mu X.\phi} = g_\gamma$  with  $g_0(v)=0$ 
(and $\semek{\nu X.\phi} = g_\gamma$  with  $g_0(v)=\infty$)
for all $v \in V$ where
\begin{displaymath} g_{\alpha} = \left\{
\begin{array}{ll}
\sem{\phi}_{\varepsilon[X \leftarrow g_{\alpha-1}]} &
    \text{for } \alpha \text{ successor ordinal,}\\
\lim_{\beta < \alpha}\sem{\phi}_{\varepsilon[X \leftarrow g_{\beta}]} &
    \text{for } \alpha \text{ limit ordinal,}
\end{array} \right.
\end{displaymath}
and $\gamma$ is such that $g_\gamma = g_{\gamma+1}$.
\end{proposition}

\vskip -0.4cm
\section{Negation and Duality}
So far, the quantitative logics $\Qmu$ and $\QML$ lack 
a negation operator and the associated 
dualities between $\land$ and $\lor$,
$\Diamond$ and $\Box$, and between least and greatest
fixed points.
Let us clarify in the following definition what we expect from such an operator.

\begin{definition} \label{defneg}
A \emph{negation operator} $f_\lnot$ for \Qmu is a function
$\Rplusinf \to \Rplusinf$, such that when we define 
$\sem{\lnot \phi} = f_{\lnot}(\sem{\phi})$,
the following equivalences hold for every $\phi \in \Qmu$: 
\begin{enumerate}
\item $\lnot \lnot \phi \equiv \phi$
\item $\lnot (\phi \land \psi) \equiv \lnot \phi \lor \lnot \psi$ and
      $\lnot (\phi \lor \psi) \equiv \lnot \phi \land \lnot \psi$
\item $\lnot \Box \phi \equiv \Diam \lnot \phi$ and
      $\lnot \Diam \phi \equiv \Box \lnot \phi$
\item $\lnot d \cdot \phi \equiv \beta(d) \cdot \lnot \phi$
         for some $\beta$ independent of $\phi$
\item $\lnot \mu X. \phi \equiv \nu X. \lnot \phi [X / \lnot X]$ and
      $\lnot \nu X. \phi \equiv \mu X. \lnot \phi [X / \lnot X]$
\end{enumerate}
\end{definition}

A straightforward calculation 
shows that the function

\begin{displaymath}
f_{\frac{1}{x}}: \Rplusinf \to \Rplusinf : x \mapsto \left\{
\begin{array}{ll} 
1/x & \text{ for } x \neq 0, x \neq \infty, \\
\infty & \text{ for } x = 0, \\
0 & \text{ for } x = \infty,
\end{array} \right. 
\end{displaymath}
is a negation operator for $\Qmu$.
Hence, we can safely include negation into the definition of $\Qmu$.
If we do so, we of course have to demand that the fixed-point variables
in the definition of least and greatest fixed point formulae,
see Definition \ref{fp}, only occur under an even number of negations,
so as to preserve monotonicity.

Moreover, 
we show that $f_{\frac{1}{x}}$ is
the only negation operator with the required properties.
You should note that this is the case even for non-discounted transition
systems, and thus it motivates our definition of the semantics of \Qmu,
in particular of the modal operators, on quantitative transition systems.

\begin{theorem}\label{onlynegation}
$f_{\frac{1}{x}}$ is the only negation operator for $\Qmu$,
even for non-discounted systems.
\end{theorem}

\vskip -0.3cm
\section{Quantitative Parity Games}

Quantitative parity games are an extension of classical parity games. 
The two main differences are the possibility
to assign real values in final positions to denote the payoff for \pzero and
the possibility to discount payoff values on edges.

\begin{definition}
A \emph{quantitative parity game} is a tuple $\calG = (V, V_0, V_1, E, \delta, \lambda, \Omega)$
where $V$ is a disjoint union of $V_{0}$ and $V_{1}$,
\ie positions belong to either \pzero or 1.
The transition relation  $E \subseteq V \times V$ describes
possible moves in the game and 
$\delta : V \times V \to \Rplus$ maps every move to
a positive real value representing the discount factor. 
The payoff function
$\lambda : \{v \in V : vE = \emptyset\} \to \Rplusinf$ assigns values to
all terminal positions and the priority function
$\Omega : V \to \{0, \ldots, n \}$ assigns a priority to every position.
\end{definition}

\noindent \textbf{How to play.}
Every play starts at some vertex $v \in V$.
For every vertex in $V_{i}$, Player $i$  chooses a successor vertex,
and the play proceeds from that vertex.
If the play reaches a terminal vertex, it ends.
We denote by $\pi = v_0 v_1 \ldots$ the (possibly infinite) play through
vertices $v_0 v_1 \ldots$, given that $(v_n, v_{n+1}) \in E$ for every $n$.
The outcome $p(\pi)$ of a finite play $\pi= v_0 \ldots v_k$ can be computed
by multiplying all discount factors seen throughout the play with the value
of the final node, 
\[p(v_0 v_1 \ldots v_k) = \delta(v_0, v_1) \cdot \delta (v_1, v_2) \cdot \ldots
\cdot \delta(v_{k-1}, v_k) \cdot \lambda(v_k).\]
The outcome of an infinite play depends only on the lowest priority seen 
infinitely often. We will assign the value $0$ to every infinite play,
where the lowest priority seen infinitely often is odd,
and $\infty$ to those, where it is even.

\noindent\textbf{Goals.}
The two players have opposing objectives regarding the outcome of the play.
\pzero wants to maximise the outcome, while \pone wants to minimise it.

\noindent \textbf{Strategies.} 
A strategy for player $i \in {0,1}$ is a function
$s : V^*V_i \to V$ with $(v, s(v)) \in E$.
A play $\pi = v_0 v_1 \ldots$ is \emph{consistent with a strategy} $s$
for player $i$, if $v_{n+1}=s(v_0\ldots v_n)$ for every $n$ such that
$v_n\in V_i$.
For strategies $\sigma, \rho$ for the two players, 
we denote by $\pi_{\sigma, \rho}(v)$ the unique
play starting at node $v$ which is consistent with both $\sigma$ and
$\rho$.

\noindent\textbf{Determinacy.} 
A game is \emph{determined} if, for each position $v$, the highest outcome
\pzero can assure from this position and the lowest outcome \pone can assure
converge,
\[ \adjustlimits\sup_{\sigma \in \Gamma_0} \inf_{\rho \in \Gamma_1}
      p(\pi_{\sigma, \rho}(v)) = 
  \adjustlimits\inf_{\rho \in \Gamma_1} \sup_{\sigma \in \Gamma_0}
      p(\pi_{\sigma, \rho}(v)) =: \val \calG (v),\]
where $\Gamma_0, \Gamma_1$ are the sets of all possible strategies 
for \pzero, \pone
and the achieved outcome is called the \emph{value of $\calG$ at $v$}.

Classical parity games can be seen as a special case of quantitative parity
games when we map winning to payoff $\infty$ and losing to payoff $0$.
Formally, we say that a quantitative parity game
$\calG = (V, V_0, V_1, E, \delta, \lambda, \Omega)$
is \emph{qualitative} when $\lambda(v) = 0$ or $\lambda(v) = \infty$ for
all $v \in V$ with $vE = \emptyset$.
In qualitative games, we denote by $W_i \in V$ the winning region of player $i$,
\ie $W_0$ is the region where player $0$ has a strategy to guarantee payoff
$\infty$ and $W_1$ is the region where player $1$ can guarantee payoff $0$.
Note that there is no need for the discount function $\delta$ in
the qualitative case as the payoff can not be changed by discounting.

Qualitative parity games have been extensively studied in the past.
One of their fundamental properties is \emph{positional determinacy}.
In every parity game, the set of positions
can be partitioned into the winning regions $W_0$ and $W_1$
for the two players,
and each player has a positional winning strategy on her winning
region (which means that the moves selected by the strategy only
depend on the current position, not on the history of the play).

Unfortunately, this result does not generalise to quantitative parity games. 
Example \ref{bsp1} shows that there are simple quantitative games
where no player has a positional winning strategy. In the depicted game
there is no optimal strategy for \pzero, and even if one fixes an approximation
of the game value, \pzero needs infinite memory to reach this approximation,
because she needs to loop in the second position as long as \pone looped in
the first one to make up for the discounts. 
(By convention, we depict positions
of \pzero with a circle and of \pone with a square and the number inside is
the priority for non-terminal positions and the payoff in terminal ones.)

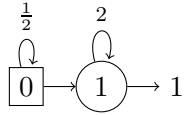
\begin{example} \label{bsp1}
~\\
\begin{center}
\vskip -0.7cm
\begin{tikzpicture}
\node [draw] (a) at (1,0) {$0$};
\node [circle,draw] (b) at (2,0) {$1$};
\node [] (c) at (3,0) {$1$};

\path[->] (a) edge [loop above] node[above] {$\scriptstyle{\frac{1}{2}}$} (a);
\path[->] (a) edge node[above] {} (b);
\path[->] (b) edge [loop above] node {$\scriptstyle{2}$} (b);
\path[->] (b) edge node[above] {} (c);
\end{tikzpicture} 
\end{center}
\end{example}

\vskip -0.4cm
\subsection{Model Checking Games for \Qmu}

A game $(\calG, v)$ is a model checking game for a formula $\phi$ and
a structure $\calK, v'$, if the value of the game starting from $v$ is
exactly the value of the formula evaluated on $\calK$ at $v'$.
In the qualitative case, that means, that $\phi$ holds in $\calK, v'$ if
\pzero wins in $\calG$ from~$v$.

\begin{definition}
For a quantitative transition system $\calK = (S, T, \delta_S, P_i)$ and
a \Qmu-formula $\phi$ in negation normal form, the quantitative parity game 
$\mc[\calK,\phi] = (V, V_0, V_1, E, \delta, \lambda, \Omega)$,
which we call the \emph{model checking game} for $\calK$ and $\phi$,
is constructed in the following way.
 
\noindent\textbf{Positions.}
The positions of the game are the pairs $(\psi, s)$,
where $\psi$ is a subformula of $\phi$, and
$s \in S$ is a state of the QTS $\calK$, 
and the two special positions $(0)$ and $(\infty)$.
Positions $(\psi,s)$ where the top operator
of $\psi$ is $\Box, \land$, or $\nu$ belong to \pone and
all other positions belong to \pzero.

\noindent \textbf{Moves.} 
Positions of the form $(|P_i-c|, s), (0),$ and $(\infty)$
are terminal positions.
From positions of the form $(\psi \land \theta, s)$,
resp. $(\psi \lor \theta, s)$, one can move to 
$(\psi, s)$ or to  $(\theta, s)$.
Positions of the form $(\Diam \psi, s)$ have
either a single successor $(0)$, in case $s$ is a terminal 
state in $\calK$, or one successor $(\psi, s')$ for every $s' \in sT$.
Analogously, positions of the form $(\Box \psi, s)$ have 
a single successor $(\infty)$, if $sT=\emptyset$,
or one successor $(\psi, s')$ for every $s' \in sT$ otherwise.
Positions of the form $(d \cdot \psi, s)$ have a unique successor
$(\psi, s')$.
Fixed-point positions $(\mu X. \psi, s)$, resp. $(\nu X. \psi, s)$
have a single successor $(\psi, s)$.
Whenever one encounters a position where the fixed-point variable stands alone,
\ie $(X, s')$, the play goes back to the corresponding definition,
namely $(\psi, s')$.

\noindent \textbf{Discounts.}
The discount of an edge is $d$ for transitions from positions
$(d \cdot \psi, s)$, it is
$\delta_S(s,s')$ for transitions from $(\Diam \psi, s)$
to $(\psi, s')$, it is $1 / \delta_S(s,s')$ for transitions from
$(\Box \psi, s)$ to $(\psi, s')$, and $1$ for
all outgoing transitions from other positions.

\noindent \textbf{Payoffs.} 
The payoff function $\lambda$ assigns $|\sem {P_i}(s)-c|$
to all positions $(|P_i-c|, s)$, $\infty$ to position $(\infty)$,
and $0$ to position $(0)$.

\noindent \textbf{Priorities.} 
The priority function $\Omega$ is defined as in the classical case
using the alternation level of the fixed-point variables, 
see \eg \cite{Graedel03}. Positions $(X, s)$ get a lower priority
than positions $(X', s')$ if $X$ has a lower alternation level than $X'$.
The priorities are then adjusted to have the right parity,  so that
an even value is assigned to all positions $(X, s)$ where $X$ is a 
$\nu$-variable and an odd value to those where $X$ is a $\mu$-variable.
The maximum priority, equal to the alternation depth of the formula,
is assigned to all other positions.

\end{definition}

It is well-known that qualitative parity games are model checking
games for the classical \muc, see e.g. \cite{EmersonJS93} or \cite{Stirling96}.
A proof that uses the unfolding technique can be found in \cite{Graedel03}.
We generalise this connection to the quantitative setting as follows.

\begin{theorem} \label{mccorrect}
For every formula $\phi$ in \Qmu, a quantitative transition system $\calK$,
and $v \in \calK$, the game $\mc[\calK,\phi]$ is determined and
\[\val \mc[\calK,\phi](\phi, v) = {\sem \phi}^\calK(v).\]
\end{theorem}

\vskip -0.3cm
\subsection{Unfolding Quantitative Parity Games}

To prove the model checking theorem in the quantitative case,
we start with games with one priority, known as reachability and safety games.
The construction of $\varepsilon$-optimal strategies is obtained
by a generalisation of backwards induction. 
At first, we fix the notation and show a few basic properties.

\begin{definition}
A number $k \in \Rplusinf$ is called \emph{$\varepsilon$-close} to
$p \in \Rplusinf$, when either $p$ is finite and $|k-p| \leq \varepsilon$ or
$p = \infty$ and $k \geq \frac{1}{\varepsilon}$.
A strategy $\sigma$ in a determined game $\calG$ is
\emph{$\varepsilon$-optimal} from $v$
if it assures a payoff $\varepsilon$-close to $\val\calG(v)$.
Furthermore, we say that $k$ is
\emph{$\varepsilon$-above} $p$ (or \emph{$\varepsilon$-below}), if
$k \geq p'$ (or $k \leq p'$) for some $p'$ that is $\varepsilon$-close to $p$.
\end{definition}

We slightly abuse the word ``close'' as $\varepsilon$-closeness
is \emph{not} symmetric, since  
$\frac 1 \varepsilon$ is $\varepsilon$-close to $\infty$,
but $\infty$ is not $\varepsilon$-close to any number $r \in \Rplus$.
Still, the following lemmas should convince you that our definition suits
our considerations well.

\begin{definition}
For every history $h = v_0 \ldots v_\ell$ of a play, let
$\Delta(h) = \Pi_{i<\ell} \delta(v_i, v_{i+1})$ be the product of
all discount factors seen in $h$, and let
$D(h) = \max (\Delta(h), \frac 1 {\Delta(h)}).$
Note that for every play $\pi = v_0 v_1\ldots$ and every $k$,
\[ p(\pi)= \Delta(v_0 \ldots v_k) \cdot p(v_{k}v_{k+1} \ldots).\]
\end{definition}

\begin{lemma} \label{closeness}
Let $x,y \in \Rplusinf$, $\varepsilon \in (0,1)$,
$\Delta\in\Rplus\setminus\{0\}$,
and $D = \max\{\Delta, \frac 1 \Delta\}$.
\begin{enumerate}
\item If $x$ is $\varepsilon/D$-close to $y$,
then $\Delta \cdot x$ is $\varepsilon$-close to $\Delta \cdot y$.
This holds in particular when $\Delta = \Delta(h)$ and $D = D(h)$
for a history $h$.
\item If $x$ is $\varepsilon/2$-close to $y$ and
$y$ is $\varepsilon/2$-close to $z$, then
$x$ is $\varepsilon$-close to $z$.
\end{enumerate}
\end{lemma}

\noindent This lemma remains valid if we replace the close-relation by
the above- or below-relation.

\begin{proposition} \label{rs-det}
Reachability and Safety games are determined, for every position $v$
there exist strategies $\sigma^\varepsilon$ and $\rho^\varepsilon$ that guarantee
payoffs $\varepsilon$-above (or respectively $\varepsilon$-below)
$\val \calG (v)$.
\end{proposition}

The next step is to prove the determinacy of quantitative parity games.
For this purpose, we present a method to unfold a quantitative parity game
into a sequence of games with a smaller number of priorities.
This technique is inspired by the proof of correctness
of the model checking games for \Lmu in \cite{Graedel03}.
We can extend this method to prove Theorem \ref{mccorrect}
by showing that, as in the classical case,
the unfolding of $\mc[\calK, \phi]$ is closely related to
the inductive evaluation of fixed points in $\phi$ on $\calK$.

From now on, we assume that the minimal priority in $\calG$ is even and
call it $m$. This is no restriction, since, if the minimal priority is odd,
we can always consider the dual game, where the roles of the players
are switched and all priorities are decreased by one.
  
\begin{definition}
We define the \emph{truncated game} $\calG^- = (V, E^-, \lambda, \Omega^-)$
for a quantitative parity game $\calG = (V, E, \lambda, \Omega)$.
We assume without loss of generality that all nodes with minimal priority
in $\calG$ have unique successors with a discount of $1$.
In $\calG^-$ we remove the outgoing edge from each of these nodes.
Since these nodes are terminal positions in $\calG^-$,
their priority does not matter any more for the outcome of a play
and $\Omega^-$ assigns them a higher priority, \eg $m+1$. Formally,
\begin{align*}
 E^- &= E \setminus \{(v,v') : \Omega (v) = m \}\\
\Omega^-(v) &= \left\{
\begin{array}{ll} 
\Omega(v) & \text{ if } \Omega(v) \neq m, \\
m+1 & \text{ if } \Omega(v) = m.
\end{array} \right.
\end{align*}
The \emph{unfolding of $\calG$} is a sequence of games $\calG^-_\alpha$,
for ordinals $\alpha$, which all coincide with $\calG^-$,
except for the valuation functions $\lambda_\alpha$.
Below we give the construction of the ${\lambda_\alpha}'s$.

For all terminal nodes $v$ of the original game $\calG$
we have $\lambda_\alpha(v) = \lambda(v)$ for all $\alpha$.
For the new terminal nodes, \ie all $v \in V$, such that $vE^- = \emptyset$
and $vE =\{w\}$,  the valuation is given by:

\begin{displaymath}
\lambda_\alpha(v) = \left\{
\begin{array}{ll} 
\infty & 
    \text{ for } \alpha = 0, \\
\val \calG_{\alpha-1}^-(w) & 
    \text{ for $\alpha$ successor ordinal,}\\
\lim_{\beta < \alpha} \val \calG_{\beta}^-(w) & 
    \text{ for $\alpha$ limit ordinal.}
\end{array} \right.
\end{displaymath}

\end{definition}

The intuition behind the definition of $\lambda_\alpha$ is to give an incentive
for \pzero to reach the new terminal nodes by first giving them the best
possible valuation, and later by updating them to values of their successor
in a previous game $\calG_\beta^-, \beta < \alpha$.

To determine the value of the original game $\calG$,
we inductively compute the values for each game in $\calG_\alpha$,
until they do not change any more. 
Let $\gamma$ be an ordinal for which 
$\val \calG_{\gamma}^- = \val \calG_{\gamma+1}^-$.
Such an ordinal exists, since the values of the games in the unfolding
are monotonically decreasing (which follows from determinacy of these
games and definition).
We set $g(v) = g_\gamma(v) = \val \calG_{\gamma}^-(v)$ and show that $g$
is the value function of the original game~$\calG$.

To prove this, we need to introduce strategies for \pone and \pzero,
which are inductively constructed from the strategies in the unfolding.
To give an intuition for the construction, we view a play in $\calG$
as a play in the unfolding of $\calG$. 
Let us look more closely at the situation of each player.

\vskip -0.3cm
\subsection*{The Strategy of \pzero}
\pzero wants to achieve the value $g_\gamma(v_0)$ or
to come $\varepsilon$-close. To reach this goal,
she imagines to play in $\calG_\gamma^-$ and uses her 
$\varepsilon$-optimal strategies $\sigma_\gamma^\varepsilon$ for that game.
Between every two occurrences of nodes of minimal priority
throughout the play, she plays a strategy $\sigma_\gamma^{\varepsilon_i}$.

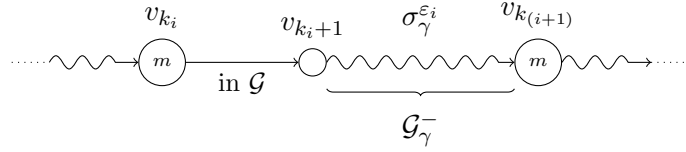
\begin{center}
\begin{tikzpicture} 
\draw (0, -1.5) node[anchor=west] 
    {\pzero's strategy after having seen $i$ nodes of priority $m$.};
\draw [dotted] (1, 0) -- (1.5, 0);
\node [draw, circle] (b) at (3,0) [label=above:$v_{k_i}$]{\tiny $m$};
\draw [->,snake=coil, segment aspect=0](1.5, 0) to node [] {}(b);
\node [draw, circle] (d) at (5,0) [label=above:$v_{k_i+1}$] {};
\path[->] (b) edge []node[below] {\small in $\calG$} (d);
\node [draw, circle] (e) at (8,0) [label=above:$v_{k_{(i+1)}}$]{\tiny $m$};
\draw [->,snake=coil, segment aspect=0](d) to node [above=6pt] 
    {$\sigma_\gamma^{\varepsilon_i}$}(e);
\draw [snake=brace, mirror snake, raise snake=10pt](d) to node [below=15pt]
    {$\calG_\gamma^-$}(e);
\draw [->,snake=coil, segment aspect=0](e) to node [] {}(9.5, 0);
\draw [dotted](9.5, 0) -- (10, 0);
\end{tikzpicture} 
\end{center}

Initially, $\varepsilon_i$ will be $\frac {\varepsilon} 2$, 
$\varepsilon$ being the approximation value she wants to attain in the end.
Then she chooses a lower  $\varepsilon_{i+1}$ 
every time she passes an edge outside of $\calG^-$.
She will adjust the approximation value not only by cutting it in half
every time she changes the strategy, but also according to
the discount factors seen so far,
since they also can dramatically alter the value of the approximation.

For a history $h$ or a full play $\pi$, let $L(h)$ (resp. $L(\pi)$)
be the number of nodes with minimal priority $m$ occurring in $h$ (or $\pi$). 
 
\begin{definition} \label{strat0}
The strategy $\sigma^\varepsilon$ for \pzero in the game $\calG$,
after history $h=v_0\dots v_\ell$ is given as follows.
In the case that $L(h)=0$ (i.e., no position of minimal priority
has been seen), let $\varepsilon':=\varepsilon/2$, and 
$\sigma^\varepsilon(h):= \sigma_\gamma^{\varepsilon'}(h)$.
Otherwise, let $v_k$ be the last node of priority $m$
in the history $h=v_0\dots v_\ell$, 
\[ \varepsilon':={\frac \varepsilon {2^{L(h)+1} D(v_0 \ldots v_k)}}.\]
and 
\[ \sigma^\varepsilon(h):= \sigma_\gamma^{\varepsilon'}(v_{k+1}\dots v_\ell).\]
\end{definition}

Now let us consider a play $\pi = v_0 \ldots v_k v_{k+1} \ldots$,
consistent with a strategy $\sigma^\varepsilon$,
where $v_k$ is the first node with minimal priority. 
The following property about values $g_\gamma(v_0)$ and $g_\gamma(v_{k+1})$
in such case (and an analogous, but more tedious one for \pone)
is the main technical point in proving $\varepsilon$-optimality.

\begin{lemma} \label{pp11}
$\Delta(v_0 \ldots v_k) \cdot g_\gamma(v_{k+1})$ is 
$\frac \varepsilon 2$-above $g_\gamma(v_0)$.
\end{lemma}

With the above lemma we prove 
the $\varepsilon$-optimality of the strategies $\sigma^\varepsilon$,
as stated in the proposition below.


\begin{proposition} \label{strat0opt}
The strategy $\sigma^\varepsilon$ is $\varepsilon$-optimal,
\ie for every $v \in V$ and every strategy $\rho$ for \pone,
$p(\pi_{\sigma^\varepsilon,\rho}(v))$ 
is $\varepsilon$-above $g(v).$
\end{proposition}

\vskip -0.4cm
\subsection*{The Strategy of  \pone} \label{playerone}
Now we look at the situation of \pone.
The problem of \pone is that he cannot just combine his strategies
for $\calG_\gamma^-$. If he did so, he would risk going infinitely often
through nodes with minimal priority which is his worst case scenario.
Intuitively speaking, he needs a way to count down,
so that will be able to come close enough to his desired value,
but will stop going through the nodes with minimal priority
after a finite number of times.
To achieve that, he utilises the strategy index as a counter.
Like \pzero, he starts with a strategy for $\calG_\gamma^-$,
but with every strategy change at the nodes of minimal priority
he not only adjusts the approximation value according to the previous one
and the discount factors seen so far,
but also lowers the strategy index in the following way.
If the current game index is a successor ordinal,
he just changes the index to its predecessor and adjusts
the approximation value in the same way \pzero does.
If the current game index is a limit value, he uses the fact,
that there is a game index belonging to a game which has an outcome
close enough to still reach his desired outcome.
In the situation depicted below he would choose an $\alpha$ such that
$\val \calG_\alpha^-(v_{k_1+1})$ is $\frac \varepsilon 4$-below
$\lambda_\gamma(v_{k_1}).$

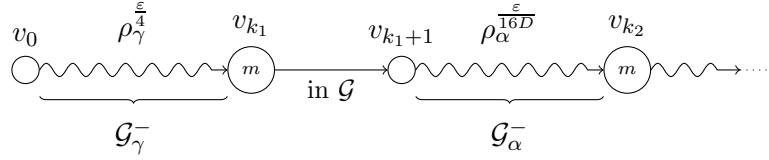
\begin{center}
\begin{tikzpicture}
\draw (-1.3, -1.5) node[anchor=west] 
    {\pone's strategy at the beginning of the play for
     a limit ordinal $\gamma$.};
\node [draw, circle] (a) at (0,0) [label=above:$v_{0}$] {};
\node [draw, circle] (b) at (3,0) [label=above:$v_{k_1}$]{\tiny $m$};
\draw [->,snake=coil, segment aspect=0](a) to node [above=6pt]
    {$\rho_\gamma^{\frac \varepsilon 4}$} (b);
\draw [snake=brace, mirror snake, raise snake=10pt](a) to node [below=15pt]
    {$\calG_\gamma^-$} (b);
\node [draw, circle] (d) at (5,0) [label=above:$v_{k_1+1}$] {};
\path[->] (b) edge []node[below] {\small in $\calG$} (d);
\node [draw, circle] (e) at (8,0) [label=above:$v_{k_2}$]{\tiny $m$};
\draw [->,snake=coil, segment aspect=0](d) to node [above=6pt]
    {$\rho_{\alpha}^{\frac \varepsilon {16 D}}$} (e);
\draw [snake=brace, mirror snake, raise snake=10pt](d) to node [below=15pt]
    {$\calG_{\alpha}^-$} (e);
\draw [->,snake=coil, segment aspect=0](e) to node [] {}(9.5, 0);
\draw [dotted](9.5, 0) -- (10, 0);
\end{tikzpicture} 
\end{center}

Finally, after a finite number of changes,  as the ordinals are well-founded,
he will be playing some version of $\rho_0^{\varepsilon_l}$ and keep
on playing this strategy for the rest of the play. 

Now we formally describe \pone's strategy.
Let us first fix some notation considering game indices.
For a limit ordinal $\alpha$, a node $v \in V$ of priority $m$,
and for $\varepsilon \in (0,1)$,  we denote by
$\alpha \restriction {\varepsilon, v}$ the index for which the value
$\val \calG^-_\alpha(v)$ 
is $\varepsilon$-below $\lambda_\alpha(w),$ where $\{w\} = vE.$

\begin{definition} \label{strat1}
For a given approximation value $\varepsilon'$, a starting ordinal $\zeta$,
and a history $h = v_0 \ldots v_l$, we define game indices
$\alpha_\zeta(h, \varepsilon')$, approximation values
$\varepsilon(h, \varepsilon')$,
and a strategy $\rho^{\varepsilon'}$ for \pone in the following way.

If $L(h) = 0$, we fix $\alpha_\zeta(h, \varepsilon') = \zeta$ and
$\varepsilon(h, \varepsilon') = \varepsilon'$.

For $h = v_0 \ldots v_k v_{k+1} \ldots v_l$,
where $v_{k}$ is the last node with minimal priority in $h$,
let $h' = v_0 \ldots v_{k-1}$ and put

\begin{displaymath}
\alpha_\zeta(h, \varepsilon') = \left\{
\begin{array}{ll} 
\alpha_\zeta(h', \varepsilon') - 1 & \text{ for } 
    \alpha_\zeta(h', \varepsilon') \text{ successor ordinal}, \\
\alpha_\zeta(h', \varepsilon') \restriction ({\frac {\varepsilon'} 
    {4^{L(h')+1} D(h')}}, v_k) & \text{ for }
    \alpha_\zeta(h', \varepsilon') \text{ limit ordinal}, \\
0 & \text{ for } \alpha_\zeta(h', \varepsilon') = 0,
\end{array} \right.
\end{displaymath}
and $\varepsilon(h, \varepsilon') = 
  {\frac {\varepsilon'} {4^{L(h)} D(v_0 \ldots v_k)}}.$

The $\varepsilon'$-optimal strategy for \pone is given by:
\[\rho_\zeta^{\varepsilon'}(v_0 \ldots v_l) = 
    \rho_{\alpha_\zeta(v_0 \ldots v_l, \varepsilon')}^
         {\frac{\varepsilon(v_0 \dots v_l, \varepsilon')}{4}}.\]
\end{definition}


\begin{proposition}\label{strat1opt}
The strategy $\rho_\zeta^\varepsilon$ is $\varepsilon$-optimal, 
\ie for every $\varepsilon \in (0,1)$, for all $v \in V$,
and strategies $\sigma$ of \pzero: $p(\pi_{\sigma,\rho_\zeta^\varepsilon}(v))$
is $\varepsilon$-below $g_\zeta(v)$. 
\end{proposition}

Having defined the $\varepsilon$-optimal strategies
$\sigma^\varepsilon$ and $\rho_\gamma^\varepsilon$,
we can formulate the conclusion.

\begin{proposition} \label{det}
For a QPG $\calG = (V, E, \lambda, \Omega)$, for all $v \in V,$
\[ \adjustlimits\sup_{\sigma \in \Gamma_0} \inf_{\rho \in \Gamma_1} p(\pi_{\sigma, \rho}(v)) =
   \adjustlimits\inf_{\rho \in \Gamma_1} \sup_{\sigma \in \Gamma_0} p(\pi_{\sigma, \rho}(v)) =
   \val \calG(v) = g(v).\]
\end{proposition}

\vskip -0.3cm
\subsection{Quantitative \muc and Games} \label{qmugames}

After establishing determinacy for quantitative parity games we are
ready to prove Theorem \ref{mccorrect}.
In the proof, we first use structural induction to show that
$\mc[\calK,\phi]$ is a model checking game for $\QML$ formulae.
Further, we only need to inductively consider formulae
of the form $\phi = \nu X.\psi$.

Note that in the game $\mc[\calQ, \phi]$, the positions with minimal priority
are of the form $(X, v)$ each with a unique successor $(\phi, v)$.
Our induction hypothesis states that for every interpretation $g$
of the fixed-point variable $X$, it holds that:
\begin{equation} \label{ind}
{\sem \phi}^\calQ_{[X \leftarrow g]} = \val \mc[\calQ, \psi[X/g]].
\end{equation}

By Theorem \ref{inductivefixpoints}, we know that we can compute
$\nu X. \psi$ inductively in the following way:
$\semek{\nu X.\psi} = g_\gamma$  with  $g_0(v)=\infty$ for all $v \in V$ and 
\[ g_{\alpha} = \left\{
\begin{array}{ll}
\sem{\psi}_{\varepsilon[X \leftarrow g_{\alpha-1}]} &
    \text{for } \alpha \text{ successor ordinal,}\\
\lim_{\beta < \alpha}\sem{\psi}_{\varepsilon[X \leftarrow g_{\beta}]} &
    \text{for } \alpha \text{ limit ordinal,}
\end{array} \right.
\]
and where $g_\gamma = g_{\gamma+1}$.

Now we want to prove that the games $\mc[\calQ, \psi[X/g_\alpha]]$
coincide with the unfolding of $\mc[\calQ, \phi].$
We say that two games coincide if the game graph is essentially the same,
except for some additional moves where neither player has an actual choice and
there is no discount that could change the outcome.
In our case these are the moves from $\phi = \nu X.\psi$ to $\psi$, which allows
us to show the following lemma.

\begin{lemma}\label{coincide}
The games $\mc[\calQ, \psi[X/g_\alpha]]$ and $\mc[\calQ, \phi]_\alpha^-$
coincide for all $\alpha.$
\end{lemma}

From the above and Proposition \ref{det},
we conclude that the value of the game $\mc[\calQ, \phi]$ is
the limit of the values $\mc[\calQ, \phi]_\alpha^-$,
whose value functions coincide with the stages of the fixed-point evaluation
$g_\alpha$ for all $\alpha$, and thus 
\[ \val \mc[\calQ, \phi] = \val \mc[\calQ, \phi]_\gamma^- = 
   g_\gamma = {\sem \phi}^\calQ. \]
\vskip -0.4cm \qed

\vskip -0.4cm
\section{Describing Game Values in \Qmu}

Having model checking games for the \qmu is just one direction
in the relation between games and logic. The other direction
concerns the definability of the
winning regions in a game by formulae in the corresponding logic.
For the classical \muc such formulae have been constructed by Walukiewicz
and it has been shown that for any parity game of fixed priority they
define the winning region for \pzero, see \eg \cite{Graedel03}.
%
%
We extend this theorem to the quantitative case in the following way.
We represent quantitative parity games 
$(V, V_0, V_1, E, \delta_G, \lambda_G, \Omega_G)$
with priorities $\Omega(V) \in \{0, \ldots d-1\}$ 
by a quantitative transition system
$\calQ_\calG = (V, E, \delta, V_0, V_1, \Lambda, \Omega)$, where
$V_i(v) = \infty$ when $v \in V_i$ and $V_i(v) = 0$ otherwise,
$\Omega(v) = \Omega_G(v)$ when $vE \neq \emptyset$ and 
$\Omega(v)=d$ otherwise,
\[
\delta(v,w) = \left\{
\begin{array}{ll} 
\delta_G(v,w) & \text{ when } v \in V_0, \\
\frac 1 {\delta_G(v,w)} & \text{ when } v \in V_1, \\
\end{array} \right. 
\]
and payoff predicate $\Lambda(v) = \lambda_G(v)$ when $vE = \emptyset$
and $\Lambda(v)=0$ otherwise.

We then build the formula $\Win d$ and formulate the theorem
\[ 
\Win d = \nu X_0.\mu X_1. \nu X_2. \ldots \lambda X_{d-1} \bigvee_{j=0}^{d-1}
((V_0 \land P_j \land \Diam X_j) \lor (V_1 \land P_j \land \Box X_j))
\lor \Lambda,
\]
where $\lambda = \nu$ if $d$ is odd, and $\lambda = \mu$ otherwise, and
$P_i := \lnot (\mu X. (2 \cdot X \lor |\Omega - i|)).$

\begin{theorem} \label{revdir-thm}
For every $d \in \N$, the value of any quantitative parity game $\calG$
with priorities in $\{0, \ldots d-1\}$ coincides with the value of $\Win d$
on the associated transition system $\calQ_\calG$.
\end{theorem}
\vskip -0.1cm

\section{Conclusions and Future Work}
In this work, we showed how the close connection between the modal
\muc and parity games can be lifted to the quantitative setting, provided
that the quantitative extensions of the logic and the games are defined
in an appropriate manner. This is just a first step in a systematic
investigation of what connections between logic and games survive in
the quantitative setting. These investigations should as well be extended
to quantitative variants of other logics, 
in particular \LTL, \CTL, \CTLstar, and \PDL.

Following \cite{dA:metrics} we work with games where discounts are multiplied
along edges and values range over the non-negative reals with infinity.
Another natural possibility is to use addition instead of multiplication
and let the values range over the reals with $- \infty$ and $+ \infty$.
Crash games, recently introduced in \cite{Seidl07}, are defined in such a way,
but with values restricted to integers. Gawlitza and Seidl present an
algorithm for crash games over finite graphs which is based on strategy
improvement \cite{Seidl07}. It is possible to translate back and forth
between quantitative parity games and crash games with real values 
by taking logarithms of the discount values on edges as payoffs
for moves in the crash game. The exponent of the value of such a crash game
is then equal to the value of the original quantitative parity game. This
suggests that the methods from \cite{Seidl07} can be applied to quantitative
parity games as well. This could lead to efficient model-checking algorithms
for $\Qmu$ and would thus further justify the game-based approach to 
model checking modal logics.

\bibliographystyle{plain}

\vskip -0.3cm
\begin{thebibliography}{10}

\bibitem{Alfaro03}
Luca de~Alfaro.
\newblock Quantitative verification and control via the mu-calculus.
\newblock In Roberto~M. Amadio and Denis Lugiez, editors, {\em CONCUR}, volume
  2761 of {\em LNCS}, pages 102--126. Springer, 2003.

\bibitem{AlfaroFHMS05}
Luca de~Alfaro, Marco Faella, Thomas~A. Henzinger, Rupak Majumdar, and
  Mari{\"e}lle Stoelinga.
\newblock Model checking discounted temporal properties.
\newblock {\em Theoretical Computer Science}, 345(1):139--170, 2005.

\bibitem{dA:metrics}
Luca de~Alfaro, Marco Faella, and Mari{\"e}lle Stoelinga.
\newblock Linear and branching system metrics.
\newblock Technical Report ucsc-crl-05-01, School of Engineering, University of
  California, Santa Cruz, 2005.

\bibitem{AlfaroHM03}
Luca de~Alfaro, Thomas~A. Henzinger, and Rupak Majumdar.
\newblock Discounting the future in systems theory.
\newblock In Jos C.~M. Baeten, Jan~Karel Lenstra, Joachim Parrow, and
  Gerhard~J. Woeginger, editors, {\em ICALP}, volume 2719 of {\em Lecture Notes
  in Computer Science}, pages 1022--1037. Springer, 2003.

\bibitem{AlfaroM04}
Luca de~Alfaro and Rupak Majumdar.
\newblock Quantitative solution of omega-regular games.
\newblock {\em J. Comput. Syst. Sci.}, 68(2):374--397, 2004.

\bibitem{EmersonJS93}
  E.~Allen Emerson, Charanjit~S. Jutla, and A.~Prasad Sistla.
  \newblock On model-checking for fragments of $\mu$-calculus.
  \newblock In {\em CAV 93}, volume 697 of \textit{Lecture Notes in
  Computer Science}, pages 385--396.  Springer, 1993.

\bibitem{Seidl07}
  Thomas Gawlitza and Helmut Seidl.  \newblock Computing game values
  for crash games.  \newblock In Kedar S. Namjoshi \textit{et al.}, eds, {\em ATVA}, {\em Lect.  Notes in
  Comp.  Science} 4762, pp.  177-191.  Springer, 2007.

\bibitem{Hugo07}
  Hugo Gimbert and Wieslaw Zielonka.  \newblock Perfect information
  stochastic priority games.  \newblock In Lars Arge \textit{et al.}, 
  eds, {\em
  ICALP}, {\em Lect. Notes in Comp. Science} 4596,
  pp. 850-861.  Springer, 2007.

%
%
\bibitem{Graedel03}
Erich Gr{\"a}del.
\newblock Finite model theory and descriptive complexity.
\newblock In {\em Finite Model Theory and Its Applications}, pages 125--230.
  Springer-Verlag, 2007.

\bibitem{spm}
Marcin Jurdzi{\'n}ski.
\newblock Small progress measures for solving parity games.
\newblock In Horst Reichel and Sophie Tison, editors, {\em STACS}, volume 1770
  of {\em Lecture Notes in Computer Science}, pages 290--301. Springer, 2000.

\bibitem{McIver}
Annabelle McIver and Carroll Morgan.
\newblock Results on the quantitative $\mu$-calculus q{M}{$\mu$}.
\newblock {\em ACM Trans. Comput. Log.}, 8(1), 2007.

\bibitem{Stirling96}
Colin Stirling.
\newblock Games and modal mu-calculus.
\newblock In Tiziana Margaria and Bernhard Steffen, editors, {\em TACAS},
  volume 1055 of {\em Lecture Notes in Computer Science}, pages 298--312.
  Springer, 1996.

\end{thebibliography}

\end{document}